\documentclass[traditabstract]{aa} 
\usepackage{graphicx}
\usepackage{longtable}
\usepackage{rotating}
\usepackage{lscape}
\usepackage{amsmath}

\begin{document}

\title{The first massive compact companion in a wide orbit around a hot subdwarf star}

\author{S.~Geier \inst{1}
   \and M.~Dorsch \inst{1,2}
   \and H.~Dawson \inst{1}
   \and I.~Pelisoli \inst{3}
   \and J.~Munday \inst{3,4}
   \and T.~R.~Marsh \inst{3}
   \and V.~Schaffenroth \inst{1}
   \and U.~Heber \inst{2}
   }

\offprints{S.\,Geier,\\ \email{sgeier@astro.physik.uni-potsdam.de}}

\institute{Institut f\"ur Physik und Astronomie, Universit\"at Potsdam, Haus 28, Karl-Liebknecht-Str. 24/25, D-14476 Potsdam-Golm, Germany
\and Dr.~Karl~Remeis-Observatory \& ECAP, Astronomical Institute, Friedrich-Alexander University Erlangen-Nuremberg, Sternwartstr.~7, D 96049 Bamberg, Germany
\and Department of Physics, University of Warwick, Conventry CV4 7AL, UK
\and Isaac Newton Group of Telescopes, Apartado de Correos 368, E-38700 Santa Cruz de La Palma, Spain}

\date{Received \ Accepted}

\abstract{We report the discovery of the first hot subdwarf B (sdB) star with a massive compact companion in a wide ($P=892.5\pm60.2\,{\rm d}$) binary system. It was discovered based on an astrometric binary solution provided by the Gaia mission Data Release 3. We performed detailed analyses of the spectral energy distribution (SED) as well as spectroscopic follow-up observations and confirm the nature of the visible component as a sdB star. The companion is invisible despite of its high mass of  $M_{\rm comp}=1.50_{-0.45}^{+0.37}\,M_{\rm \odot}$. A main sequence star of this mass would significantly contribute to the SED and can be excluded. The companion must be a compact object, either a massive white dwarf or a neutron star. Stable Roche lobe overflow to the companion likely led to the stripping of a red giant and the formation of the sdB, the hot and exposed helium core of the giant. Based on very preliminary data, we estimate that $\sim9\%$ of the sdBs might be formed through this new channel. This binary might also be the prototype for a new progenitor class of supernovae type Ia, which has been predicted by theory.

}

\authorrunning{Geier et al.}
\titlerunning{Massive compact companion around hot subdwarf}

\maketitle

\section{Introduction \label{sec:intro}}

The Gaia mission Data Release 3 (Gaia collaboration \cite{gaia22}) opened up a new window to study binary star systems releasing several hundred thousand orbital solutions of such systems based on high-precision astrometry (Halbwachs et al. \cite{halbwachs22}). This unique dataset allows us to probe a yet unexplored parameter space of the binary population. In particular, targeted searches for the predicted population of non-interacting massive, compact companions such as massive white dwarfs (WDs), neutron stars (NSs) or black holes (BHs) can now be conducted for the first time. It is therefore not surprising, that the first candidates for such binaries have been discovered not long after Gaia DR3 had been released, among them a Sun-like main sequence star and a red giant with BH companions as highlights (Andrews et al. \cite{andrews22}; Tanikawa et al. \cite{tanikawa23}; Chakrabarti et al. \cite{chakrabarti23}; El-Badry et al. \cite{elbadry23}; Shahaf et al. \cite{shahaf23}). 

Andrews et al. (\cite{andrews22}) selected candidates for massive compact companions from Gaia DR3 by applying several quality criteria and by calculating the astrometric binary mass function as outlined in Halbwachs et al. (\cite{halbwachs22}). To determine the mass of the unseen companion, the mass of the visible star needs to be constrained. To achieve that, Andrews et al. (\cite{andrews22}) used the parameter inference system Apsis provided by the Gaia team for most of their 24 candidates. However, for eight stars they obtained follow-up spectroscopy and determined the atmospheric parameters by fitting model spectra. Surprisingly, one of those stars turned out not to be a cool main sequence (MS) star as expected. Gaia\, DR3\,3649963989549165440 (BPS\,BS\,16981$-$0016) shows a spectrum characteristic of a rare hot subluminous B star (sdB), which was confirmed by a comparison with model spectra. 

Hot subdwarf stars (sdO/Bs) are smaller and of lower mass than hot main sequence stars with those spectral types. Most of them are core helium-burning, extreme horizontal branch (EHB) stars with very thin hydrogen envelopes (Heber et al. \cite{heber86}). The formation of sdO/Bs requires that their usually low- to intermediate mass progenitors lose their envelopes almost completely towards the end of the first red-giant phase. Binary interactions are likely responsible for this mass loss (Pelisoli et al. \cite{pelisoli20}; Geier et al. \cite{geier22}). This is supported by the high binary fraction of the sdB stars. Two thirds of them are either found in close binaries ($P=0.03-27\,{\rm d}$) with faint cool or compact companions such as WDs or in wider binaries ($P=300-1200\,{\rm d}$) in orbit with F/G/K-type main-sequence stars (e.g. Barlow et al. \cite{barlow13}; Vos et al. \cite{vos18}; Copperwheat et al. \cite{copperwheat11}; Kawka et al. \cite{kawka15}; Schaffenroth et al. \cite{schaffenroth22}). While the close binaries are formed most likely by an unstable phase of mass transfer involving a common envelope, stable mass transfer to a main sequence companion and the merger of two helium white dwarfs have been proposed as possible formation channels as well  (Han et al. \cite{han02,han03}; Clausen et al. \cite{clausen12}; Zhang \& Jeffery \cite{zhang12}; Chen et al. \cite{chen13}; Xiong et al. \cite{xiong17}). 

Assuming the canonical mass ($0.47\,M_{\rm \odot}$) of a degenerate red-giant core which ignited helium burning through a flash for the sdB,  Andrews et al. (\cite{andrews22}) determine a mass of $1.41_{-0.34}^{+0.62}\,M_{\rm \odot}$ for the unseen companion of BPS\,BS\,16981$-$0016. The authors conclude that an F-type main sequence companion, which would be consistent with the derived mass, might still be outshined by the sdB and can therefore not be excluded based on their analysis. 

If confirmed, however, BPS\,BS\,16981$-$0016 would be the first sdB with such a massive unseen companion. Here we perform a detailed analysis of this system, exclude an MS companion, and confirm it to be the first of its kind.

\section{Observations and data analysis}

BPS\,BS\,16981$-$0016 has been identified as field horizontal branch star based on an objective prism spectrum by Beers et al. (\cite{beers96}) and classified as sdB star based on spectroscopy in Geier et al. (\cite{geier17b}). The parameters determined in the following analyses and taken from the literature are summarized in Table~\ref{tab:params}.

\subsection{Quantitative spectroscopic analysis}

To refine the spectroscopic classification, determine the atmospheric parameters, and search for spectroscopic signatures of a potential companion as well as binary-induced radial velocity (RV) variations, we obtained spectroscopic observations of BPS\,BS\,16981$-$0016. 

A spectrum taken with the intermediate dispersion spectrograph (IDS, grating R400B, $R\simeq1400$, $3600-5900\,{\rm \AA}$) at the Isaac Newton Telescope (INT) on 29.6.2010 has been downloaded from the ING data archive and sixteen additional spectra have been taken with IDS (grating R600R, $R\simeq3000$, $4400-6700\,{\rm \AA}$) distributed over seven nights from 5.8. to 11.8.2022. Spectra of arc lamps for wavelength calibration have been taken before or after the exposures at the position of the stars observed. These data have been reduced using the PAMELA and MOLLY packages (Marsh \cite{marsh89}) and independently also with IRAF to quantify systematic differences, which turned out to be small.

The spectra clearly show broadened hydrogen Balmer and the neutral helium lines at $4472\,{\rm \AA}$, $5876\,\,{\rm \AA}$ and $6678\,{\rm \AA}$ characteristic of an sdB star. Using the IDS spectrum from 2010 we performed a quantitative spectroscopic analysis as outlined in Irrgang et al. (\cite{irrgang14}) using a state-of-the-art hybrid LTE/NLTE model grid computed with the three codes ATLAS12 (Kurucz \cite{kurucz96}), DETAIL and SURFACE (Giddings \cite{giddings81}; Butler \& Giddings \cite{butler85}, Przybilla, Nieva \& Butler \cite{przybilla11} and Irrgang et al. \cite{irrgang18}); for the details of recent improvements of the model atmosphere codes and the fitting procedure see Irrgang et al. \cite{irrgang21,irrgang22}. The grid also takes into account the chemical peculiarities of sdB stars. To cope with the large spread of helium abundances observed for hot subdwarf stars the grid covers helium to hydrogen ratios from $10^{-5}$ to 300. The metal abundance patterns of typical sdB stars are very different from that of the Sun. While in sdB stars the lighter metals (C to Ar) are subsolar, heavier elements of the iron group have supersolar abundances (see Geier \cite{geier13a}). Iron appears to be an exception as its abundance appears to be close to solar on average. We account for this abundance pecularities by adopting the average metal composition suggested by Pereira (\cite{pereira11}, see also Naslim et al. \cite{naslim13}) to calculate atmospheric models and flux distributions. The ATLAS12 code uses the opacity sampling technique to account for line blanketing, which is flexible and allows to adopt any pre-defined abundance distribution. The telluric lines were matched simultaneously to the observations. Lines of hydrogen and neutral helium have been fitted to obtain the atmospheric parameters and the radial velocity (see Fig.~\ref{spectro_fit_2010}). 

The resulting atmospheric parameters $T_{\rm eff}=29560\pm270\,{\rm K}$, $\log{g}=5.88\pm0.05$ and $\log{n({\rm He)}/n({\rm H)}}=-2.66\pm0.13$ are typical for sdB stars. The statistical $1\sigma$ uncertainties provided here are smaller than the typical systematic uncertainties in this parameter range. Hence, we added systematic uncertainties  ($\Delta T_{\rm eff}/T_{\rm eff}\sim0.03$, $\Delta \log{g}\sim0.10$, $\Delta \log{n({\rm He)}/n({\rm H)}}\sim0.10$) in quadrature before performing the binary analysis. The spectra show none of the typical signs of a cool MS F/G/K-type companion such as the G-band, Ca\,II H/K or Mg\,I lines.
Sixteen spectra taken in 2022 spread out over 7 nights were matched with synthetic spectra to determine radial velocities (Table~\ref{tab:rvs}). There are no variations on a time scale of few days at a level of $\Delta$RV$\approx$10km\,s$^{-1}$.

\begin{figure}[t!]
\begin{center}
	\resizebox{8cm}{!}{\includegraphics{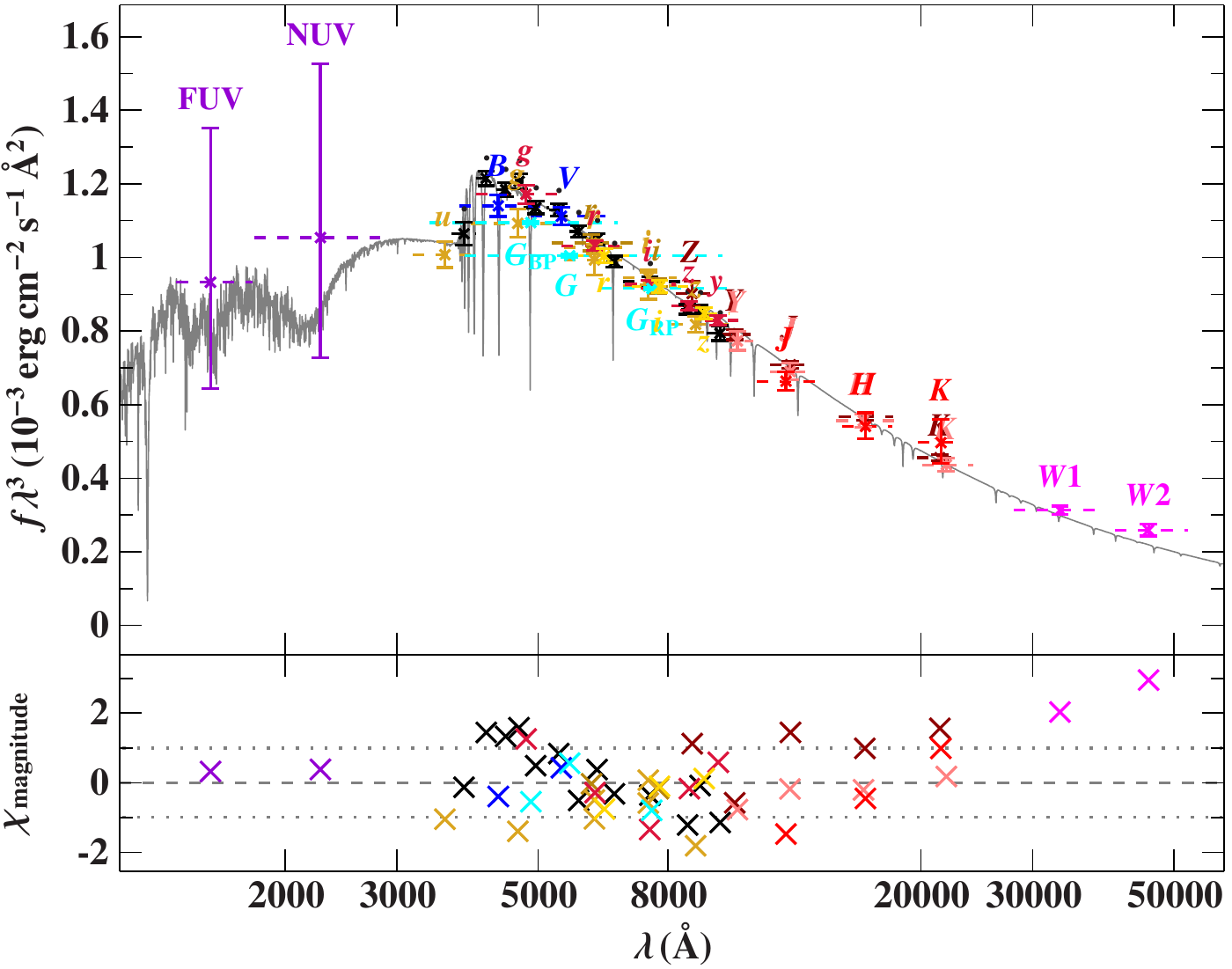}}
\end{center} 
\caption{Top panel: Spectral energy distribution of BPS\,BS\,16981$-$0016. Filter-averaged fluxes converted from observed magnitudes from different photometric surveys are shown (violet: GALEX, Bianchi et al. \cite{bianchi17}; bright yellow: DELVE, Drlica-Wagner et al. \cite{drlica21}; cyan: Gaia EDR3, Riello et al. \cite{riello21}; black: Gaia DR3 XP box filters, Gaia collaboration \cite{gaia22}; blue: APASS9, Henden et al. \cite{henden16}; yellow: SDSS, Alam et al. \cite{alam15}; dark yellow: Skymapper, Onken et al. \cite{onken19}; pink: UKIDSS, Lawrence et al. \cite{lawrence07}; dark red: VIKING, Edge et al. \cite{edge13}; red: 2MASS, Skrutskie et al. \cite{skrutskie06}; magenta: WISE, Schlafly et al. \cite{schlafly19}). The respective full width at tenth maximum are shown as dashed horizontal lines. The best-fitting model, degraded to a spectral resolution of $6\,{\rm \AA}$ is plotted in gray. In order to reduce the steep SED slope the flux is multiplied by the wavelength cubed. Bottom panel: Difference between synthetic and observed magnitudes divided by the corresponding uncertainties (residual $\chi$).}
\label{sed}
\end{figure}

\subsection{Spectral energy distribution}

To search for an infrared (IR)-excess indicative of a cooler main sequence companion and analyse the sdB primary, we fitted the spectral energy distribution (SED) of BPS\,BS\,16981$-$0016 to a synthetic flux distribution calculated with the same model atmosphere grid we used for the spectroscopic analysis. This method is described in more detail by Heber et al. (\cite{heber18}). Photometric data ranging from the ultraviolet to the infrared (IR) has been gathered (see Fig.~\ref{sed}). 

The SED is perfectly fitted with a model of a single sdB with the atmospheric parameters fixed to the spectroscopically determined values. No IR-excess indicative of a cool MS companion was detected. Since this method is very sensitive, we can exclude any MS companion other than M-type in this way.   

Furthermore, we constrained the angular diameter $\log{\Theta}=-11.049_{-0.010}^{+0.011}$ and the interstellar reddening parameter $E(44-55)=0.085\pm0.002\,{\rm mag}$. Combining the angular diameter with the parallax measurement taken from the Gaia DR3 non-single stars catalogue, where the correction for astrometric motion of the binary has been applied, we constrained the radius of the visible star to $R_{\rm sdB}=0.145\pm0.007\,R_{\rm \odot}$ again typical for an sdB. The radius and the $T_{\rm eff}$ from the spectroscopic analysis also allow us to calculate a quite typical luminosity of $14.5_{-2.1}^{+2.4}\,L_{\rm \odot}$.

Finally, the mass of the sdB can be calculated from the spectroscopically determined surface gravity, the parallax, and the angular diameter using $M_{\rm sdB}=g\Theta^{2}/4G\varpi^{2}$. The resulting mass of $0.58_{-0.15}^{+0.20}\,M_{\rm \odot}$ is somewhat higher, but within the uncertainties consistent with the canonical mass of $\sim0.47\,M_{\rm \odot}$, which is characteristic for the majority of the observed sdB stars (Vos et al. \cite{vos19}; Schaffenroth et al. \cite{schaffenroth22}; Schneider \cite{schneider22}).

\subsection{Binary analysis}\label{sec:binary}

The relevant astrometric binary parameters of BPS\,BS\,16981$-$0016, which are the orbital period $P=892.5\pm60.2\,{\rm d}$, the eccentricity of the orbit $e=0.36\pm0.14$, the orbital separation $a=2.269\pm0.056\,{\rm mas}$, and the inclination angle of the orbital plane $i=56.8\pm3.2^{\circ}$ are taken from Gaia DR3 (Gaia collaboration \cite{gaia22}). The values and uncertainties of $a$ and $i$ have been calculated by the Strasbourg astronomical Data Center (CDS) following the recipe by Halbwachs et al. (\cite{halbwachs22}) and added to the Vizier online catalogue (Gaia DR3 catalogue Part 3: Non-single stars, I/357). 

Since we have shown that there is no second component visible in this binary, we conclude that the photocenter measured by Gaia follows the visible sdB star. The astrometric mass function can now be calculated as described in Andrews et al. (\cite{andrews22}):
\begin{equation*}
\frac{M_{\rm comp}^{3}}{(M_{\rm sdB}+M_{\rm comp})^{2}}=\left(\frac{a}{\rm 1\,mas}\right)^{3}\,\left(\frac{\varpi}{\rm 1\,mas}\right)^{-3}\,\left(\frac{P}{\rm 1\,yr}\right)^{-2}\,M_{\rm \odot}
\end{equation*}
The resulting mass function $0.78\pm0.15\,M_{\rm \odot}$ together with the mass of the sdB was then used to numerically calculate the mass of the unseen companion $M_{\rm comp}=1.50_{-0.45}^{+0.37}\,M_{\rm \odot}$. This value is consistent with the one from Andrews et al. (\cite{andrews22}), who provide $2\sigma$ uncertainties in contrast to the $1\sigma$ uncertainties provided here. 

\begin{figure}[t!]
\begin{center}
	\resizebox{8cm}{!}{\includegraphics{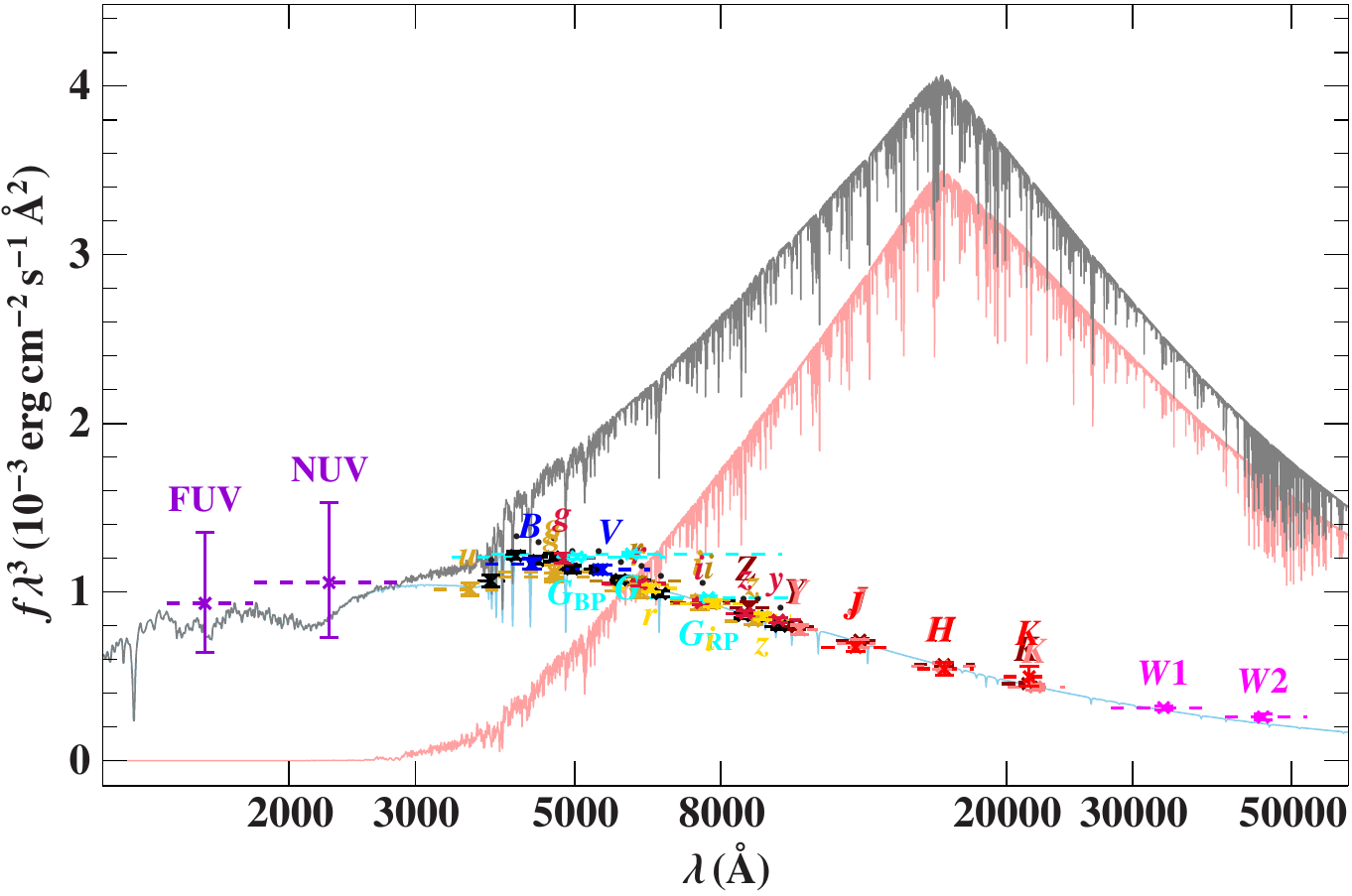}}
\end{center} 
\caption{Spectral energy distribution similar to Fig.~\ref{sed}. To illustrate the contribution of a G-type star with $1\,M_{\rm \odot}$ to the SED a model (light red) has been added to the best fit model of the sdB (light blue). The resulting composite model is shown in gray.}
\label{sed_composite}
\end{figure}

In Fig.~\ref{sed_composite} we show the SED of our best-fit sdB model combined with the model SED of a G-type star with $1\,M_{\rm \odot}$, which is the lower border of our $1\sigma$ confidence interval for the companion mass of BPS\,BS\,16981$-$0016. An MS companion with a comparable or even higher mass up to our upper border of the confidence interval of $1.9\,M_{\rm \odot}$ (corresponding to spectral type A) would significantly contribute to the optical flux and completely dominate in the IR. The single-star SED and the single-lined spectrum are therefore only consistent with a compact companion in this mass range. This could either be a very massive WD or a NS. 

An independent way to check whether the astrometric solution is correct, is provided by the RV-variability of the system. Although we do not have phase-resolved spectroscopy covering the rather long orbital period of the system, we have obtained spectra with a timebase of several days in 2022 and more than 12 years in between 2010 and 2022. 

Measuring the individual RVs along with the atmospheric parameters, we do not detect a significant shift in between the 16 spectra taken in 2022 and measure a mean RV of $-18\pm8\,{\rm km\,s^{-1}}$. However, for the 2010 spectrum we measure a quite different RV of $-107\pm7\,{\rm km\,s^{-1}}$.

To check for systematic RV shifts, which are detected regularly in medium- or low-resolution spectra, we measured the RVs of calibration stars, which have been observed along with our target in the same observing nights and reduced in the same way as described above. For the 2010 spectrum we used the cool MS star HD\,196850, which has been observed as standard during the run. We fitted the spectrum in the same way as described above using an appropriate model grid for such cool stars as described in Dorsch et al. (\cite{dorsch21}) and found a systematic shift of $-38\pm1\,{\rm km\,s^{-1}}$ between IDS and the very precise RV measured by Gaia (Gaia collaboration \cite{gaia22}). 

To check the systematic accuracy of the higher-dispersion set of IDS spectra taken in 2022, we used the two cool MS standard stars BD+16\,59 and BD+14\,323. Those have very precise and consistent multi-epoch RVs both from Gaia and the Apache Point Observatory Galactic Evolution Experiment (APOGEE, Jonsson et al. \cite{jonsson20}). We fitted two IDS spectra of BD+14\,323 and one of BD+16\,59 as described above. The average systematic shift was $+3\pm1\,{\rm km\,s^{-1}}$. 

After correcting the RV values for the systematic shifts (the corrected individual RVs are provided in Table~\ref{tab:rvs}), we measure an RV shift of $\Delta RV=48\pm11\,{\rm km\,s^{-1}}$ between 2010 and 2022. We can compare this shift with the theoretically predicted RV semi-amplitude $K$ for this binary, which can be easily calculated with the binary parameters we have already determined:
\begin{equation*}
K=\left(\frac{2\pi\,G\,M_{\rm comp}^{3}\sin{i}^3}{(M_{\rm sdB}+M_{\rm comp})^2\,P\,(1-e^{2})^{3/2}}\right)^{1/3}
\end{equation*}
The resulting $K=23\pm20\,{\rm km\,s^{-1}}$ translates into a maximum RV shift $\Delta RV_{\rm max}=46\pm20\,{\rm km\,s^{-1}}$, which matches the $\Delta RV$ measured from the spectra. The measured RV variation is therefore consistent with the astrometric solution. 

\section{The nature of the binary system and the unseen companion}

There is a known population of wide sdB binaries with F/G/K-type main sequence companions, periods of hundreds of days and small or moderate eccentricities, which are predicted to have similar astrometric parameters as BPS\,BS\,16981$-$0016. Spectroscopic binary solutions have been obtained for $\sim30$ of them so far (Vos et al. \cite{vos18}; Nemeth et al. \cite{nemeth21}; Dorsch et al. \cite{dorsch21}; Molina et al. \cite{molina22}). Most recently, Lei et al. (\cite{lei23}) indeed confirmed two of the astrometric binaries from Gaia DR3 to belong to this class. Those binaries have been predicted by binary evolution theory as the outcome of stable Roche Lobe overflow (RLOF) before (Han et al. \cite{han02,han03}) and their main properties match the theoretical predictions reasonably well (Chen et al. \cite{chen13}; Vos et al. \cite{vos20}). About one third of the known sdB population shows composite spectra or SEDs indicative of an sdB and a F/G/K-type companion, and all of them are likely the outcome of such a binary interaction (Pelisoli et al. \cite{pelisoli20}). However, the standard formation scenarios for sdBs do not predict a significant population of compact companions in wide orbits, because this would require massive and therefore rare companions (Han et al. \cite{han03}). 

Recently, Otani et al. (\cite{otani22}) reported the detection of pulsation time variations likely caused by an unseen companion with a minimum mass of $\sim0.5\,M_{\rm \odot}$ orbiting the pulsating sdB AQ\,Col with a period of $486\,{\rm d}$. The blue colour of the system indicates a compact WD companion. Based on RV-variations detected in time-resolved spectra the authors propose this object to be an hierarchical triple system, in which case the mass of the wide companion might be even higher. Better spectroscopy is needed to constrain the parameters of this interesting system.

The compact companion of BPS\,BS\,16981$-$0016 is actually among the most massive ones ever discovered orbiting a hot subdwarf. Only few of them are orbited by WDs with masses of $0.7-1.0\,M_{\rm \odot}$, all of them in very close binaries with orbital periods of less than $\sim2\,{\rm hr}$ (Maxted et al. \cite{maxted00}; Geier et al. \cite{geier07,geier13b}; Vennes et al. \cite{vennes12}; Pelisoli et al. \cite{pelisoli21}; Kupfer et al. \cite{kupfer22}). Only one compact companion to a hot subdwarf has a mass exceeding $1\,M_{\rm \odot}$. In the X-ray binary HD\,49798 a massive ($\sim1.2\,M_{\rm \odot}$) companion orbits a He-sdO companion of intermediate mass ($\sim1.5\,M_{\rm \odot}$, Mereghetti et al. \cite{mereghetti09}) with a short period of $\sim1.55\,{\rm d}$. The nature of the companion is debated. Although the X-ray properties pointed towards a neutron star (Mereghetti et al. \cite{mereghetti16}), a young and contracting WD turned out to be even more consistent with the observations (Popov et al. \cite{popov18}; Mereghetti et al. \cite{mereghetti21}). 

Candidate close binary systems with compact companions of even higher masses have been identified by Geier et al. (\cite{geier10}). The mass estimate for those companions, however, rests on the assumption that the rotation of the sdB primaries is tidally locked to their orbits. However, this assumption turned out not to be valid in all cases (Preece et al. \cite{preece18}; Schaffenroth et al. \cite{schaffenroth21}). The project ``Massive Unseen Companions to Hot Faint Underluminous Stars from SDSS" (MUCHFUSS) aimed to find hot subdwarf stars with massive compact companions (Geier et al. \cite{geier11}). From the non-detection of those systems the authors concluded that their fraction among the close binary sdBs is either smaller than $1.5\%$ or they must have orbital periods exceeding $8\,{\rm d}$ (Geier et al. \cite{geier15,geier17a}). 

All of those confirmed or candidate sdB binaries with massive compact companions are in close orbits and are therefore the outcome of unstable mass transfer and a common envelope phase (Han et al. \cite{han02,han03}). The wide orbit of BPS\,BS\,16981$-$0016 and the stripped nature of its visible sdB component makes it unique and the prototype for an entirely new class of binary systems. It also calls for a specific formation scenario. In the following we discuss several possible scenarios to explain the formation of BPS\,BS\,16981$-$0016, which are closely connected to the nature of its massive unseen companion.

\subsection{Binary with neutron star companion}

The companion mass of BPS\,BS\,16981$-$0016 is quite typical for neutron stars observed in both interacting and non-interacting binary systems. Wu et al. (\cite{wu18}) investigated the formation of sdB+NS binaries systematically with detailed binary evolution calculations. They predict such systems to form not just after a CE-phase, but also as the result of stable RLOF. Depending on the evolutionary phase at which the mass transfer starts, the orbital periods of those post-RLOF binaries can range from a few to more than $1000$ days. Since comparatively massive donor stars can start mass transfer already well before the tip of the RGB, the final orbital periods are predicted to be shortest for the most massive donors. 

\begin{figure*}[t!]
\begin{center}
	\resizebox{14cm}{!}{\includegraphics{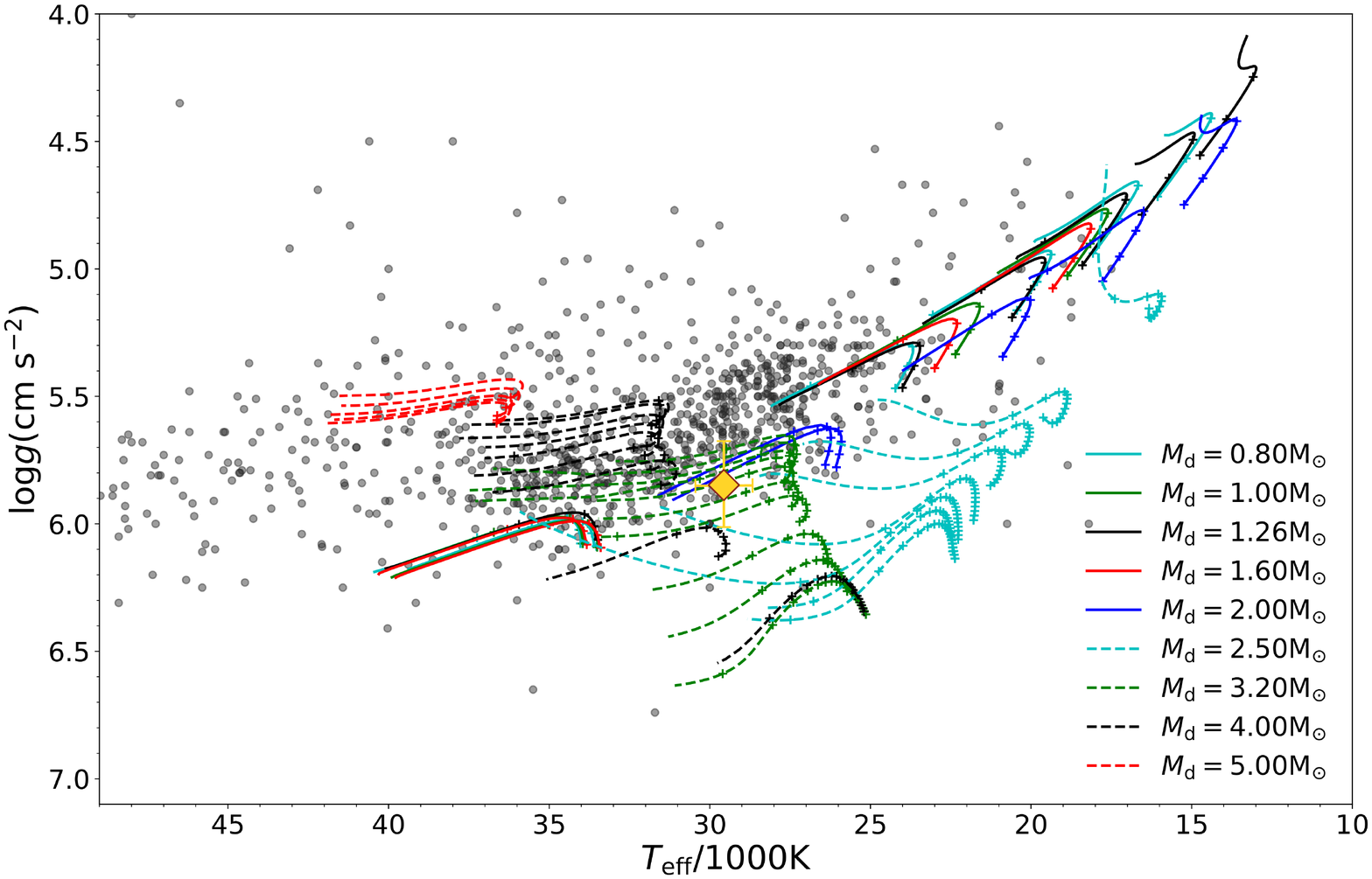}}
\end{center} 
\caption{$T_{\rm eff}-\log{g}$-diagram taken from Wu et al. (\cite{wu18}) with BPS\,BS\,16981$-$0016 marked a yellow diamond. Evolutionary tracks for donor stars with different initial masses on the main sequence (colour coded) are plotted during the He-core burning phase along with a atmospheric parameters of a sample of known hot subdwarfs taken from Geier et al. (\cite{geier17b}). The solid lines show tracks for He cores, which were degenerate before their ignition ($M_{\rm d}\leq2.0\,M_{\rm \odot}$), and the dashed lines show those from non-degenerate He cores ($M_{\rm d}\geq2.5\,M_{\rm \odot}$).}
\label{tefflogg}
\end{figure*}

Wu et al. (\cite{wu18}) also provide detailed models for initial MS donor masses $M_{\rm d}$ ranging from $0.8\,M_{\rm \odot}$ to $5.0\,M_{\rm \odot}$ and an initital NS mass of $1.4\,M_{\rm \odot}$, which can be directly compared to the observed parameters of BPS\,BS\,16981$-$0016 (see Fig.~\ref{tefflogg}, which was taken from Wu et al. \cite{wu18} and modified just by adding BPS\,BS\,16981$-$0016). For an MS donor mass of $2.0\,M_{\rm \odot}$ the predicted final periods range from $720\,{\rm d}$ to $1015\,{\rm d}$ and the radial-velocity semi-amplitudes from $20\,{\rm km\,s^{-1}}$ to $23\,{\rm km\,s^{-1}}$ consistent with the orbital parameters of BPS\,BS\,16981$-$0016. Also the atmospheric parameters of the sdB match the evolutionary tracks for the $2.0\,M_{\rm \odot}$ models. Even the predicted mass of the sdB, which ranges from $0.42\,M_{\rm \odot}$ to $0.45\,M_{\rm \odot}$ is consistent with the one we determined for the sdB in BPS\,BS\,16981$-$0016 within the uncertainties.

The models of Wu et al. (\cite{wu18}) match the observed parameters of BPS\,BS\,16981$-$0016 remarkably well and also put constraints on the formation of the sdB. Only stars with degenerate cores, which start He-burning with a He-flash at the tip of the RGB will evolve to such wide binaries. The final orbital periods of even slightly more massive donor stars are predicted to be significantly shorter ($1.5-130\,{\rm d}$), because He-burning is ignited non-degenerately. 

\subsection{Binary with massive white dwarf companion}

The mass range of the unseen companion in BPS\,BS\,16981$-$0016 would also allow it to be a massive WD.  In this case, this binary might be closely linked to an open question of high impact to astrophysics in general: the missing population of progenitor systems of supernovae type Ia (SN\,Ia). SN Ia are not only the most important rungs in the cosmic distance ladder on the largest scales, but also the main sources of iron in the universe, and they play an important role for the supernova feedback in galaxies. Despite their importance, the dominant progenitor systems of those explosions could not be clearly identified yet (Maoz et al. \cite{maoz14}; Livio \& Mazzali \cite{livio18}). 

It is consensus that the thermonuclear explosion of a WD is most consistent with the observables. But since such an explosion needs to be triggered by mass transfer from a companion star, it remains challenging that no traces of either hydrogen- or helium-rich material from such a companion can be found in the spectra of SN Ia. This is one of the reasons, why the merger of two WDs is currently more favored than mass-transfer from a normal companion star (single-degnerate scenario, SD; e.g. Tucker et al. \cite{tucker20}). However, Justham (\cite{justham11}) and Di Stefano et al. (\cite{distefano11}) proposed a peculiar SD scenario which naturally explains the lack of H- and He-rich material.                  
                          
In this scenario, a massive WD accretes material from a red giant star and grows to the Chandrasekhar mass limit. At the same time, angular momentum is transferred as well and the centrifugal force of the fast-spinning WD keeps it stable even slightly above the limiting mass. To collapse and ignite explosive carbon burning the WD has to spin down first, which happens on timescales long enough for the donor star to lose its hydrogen envelope, evolve away from the red giant branch, and become compact. The resulting SN then happens in a H- and He-free environment as expected for type Ia. 

The immediate progenitors of this scenario have been proposed to be He-WDs or hot subdwarfs with massive WDs in wide orbits of several hundred days around massive WDs (Justham \cite{justham11}; di Stefano et al. \cite{distefano11}; Benvenuto et al. \cite{benvenuto15}), consistent with the properties of BPS\,BS\,16981$-$0016. More detailed theoretical models are needed to be compared to the observations and explore this scenario further.

\subsection{Hierarchical triple system with massive outer component}

The unseen companion might also not be directly responsible for the formation of the sdB in BPS\,BS\,16981$-$0016. Triple star scenarios have been proposed to explain the formation of sdBs in triple, but also in binary systems (Preece et al. \cite{preece22}). Since Gaia can only resolve wide binaries astrometrically, BPS\,BS\,16981$-$0016 might be a hierarchical triple with a post-CE inner binary orbited by the detected outer companion. Such systems are known and all but one candidate (Otani et al. \cite{otani22}) show composite spectra indicative of wide F/G-companions, as well as high RV-shifts caused by an unseen, closer companion (Heber et al. \cite{heber02}; Barlow et al. \cite{barlow14}; Kupfer et al. \cite{kupfer15}).

A massive compact close companion such as a WD can be very likely excluded based on the lack of significant RV variations on a timescale of days within our observing run in 2022 (see Sect.~\ref{sec:binary}). A low-mass MS or brown dwarf companion, which might cause only small RV variations of the sdB on the other hand would cause a characteristic and easily detectable variation in the light curve due to irradiation of the cool companion by the hot primary. 

We have checked archival light curves of BPS\,BS\,16981$-$0016 obtained by the  Transiting Exoplanet Survey Satellite (TESS, Ricker et al. \cite{ricker15}) and the Zwicky Transient Facility (ZTF, Bellm et al. \cite{bellm19}) in the way described in Schaffenroth et al. (\cite{schaffenroth22}) for such a reflection effect and did not detect any variation (see Fig.~\ref{lc}). 

The third option would be the formation of the sdB via the merger of an inner close binary consisting of a He-WD and another WD or a low-mass MS-star. The influence of the outer companion might even induce such a merger via Zeipel-Lidov-Kozai oscillations (Preece et al. \cite{preece22}). He-WD mergers are very likely resulting in the formation of He-rich sdO/Bs (Zhang \& Jeffery \cite{zhang12}), whereas BPS\,BS\,16981$-$0016 has a quite low He-abundance. The merger of a He-WD with a low-mass MS star or substellar object might still be possible (Zhang et al. \cite{zhang17}).

Another indication that the wide companion might not have been involved in the mass-transfer leading to the formation of the sdB is the relatively high eccentricity ($e=0.36\pm0.14$) of the orbit. Mass transfer is expected to be very efficient in circularising the orbit, whereas wide non-interacting binaries show a diverse range of eccentricities (Raghavan et al. \cite{raghavan10}). However, also the known post-RLOF sdB systems with MS companions have unexpectedly high eccentricities of up to $\sim0.2$, which is likely caused by eccentricity pumping mechanisms such as phase dependent RLOF or interactions with a circumbinary disc (Vos et al. \cite{vos15}; Deca et al. \cite{deca18}). Most recently, Lei et al. (\cite{lei23}) discovered an sdB+MS system based on Gaia astrometry with an even higher eccentricity of $0.50\pm0.09$. Within the uncertainties, the eccentricity of BPS\,BS\,16981$-$0016 might therefore still be consistent with a previous RLOF episode. 

\section{Fraction of wide sdB binaries with compact companions}

Current estimates for the fraction of composite sdB binaries are of the order of $30\%$ of the sdB population (e.g. Geier et al. \cite{geier17b}; Stark \& Wade \cite{stark03}) and there is evidence that all those have interacted before (Pelisoli et al. \cite{pelisoli20}). Assuming that the orbital parameters of the solved sdB+MS binaries (Vos et al. \cite{vos18}) are representative of the whole population, they should be detectable as astrometric binaries if not too distant. We can therefore make the reasonable assumption that such MS companions to sdBs are likely to be detected as astrometric binaries by Gaia.

We cross-matched the Gaia DR3 non-single star catalogue with the most recent catalogues of hot subdwarfs and candidates (Culpan et al. \cite{culpan22}). In total, we found 13 binaries with orbital solutions. Indeed, 10 of those systems ($77\pm15$\,\%\footnote{The uncertainties are for 68\,\% confidence as estimated by drawing from the corresponding binomial distribution.}) show IR excesses in the SED indicative of F/G/K-type companions. The remaining three systems (including BPS BS 16981-0016), however, have single-star SEDs indicative of compact companions. This corresponds to an unexpectedly high fraction of $23 \pm 15$\,\% amongst the astrometric binaries.

This means that for every 100 binaries detectable by SED fitting, there seem to be an additional $30\pm16$ astrometric binaries with hidden companions. If we assume a composite-SED fraction of $30\pm5$\,\% for the hot subdwarf population, hidden astrometric WD/NS companions should therefore exist for $8.6^{+5.7}_{-4.5}$\,\% of the overall hot subdwarf population. Because composite-SED systems are not expected to host WD/NS companions, a higher hidden companion fraction of $12.2^{+8.2}_{-6.5}$\,\% would be expected for non-composite hot subwdarfs. However, these numbers should be regarded as preliminary and rough estimates, because two of the sdB+WS/NS candidates still lack spectroscopic confirmation and also because the published sample of astrometric binaries is very biased due to rigid quality control (Halbwachs et al. 2022).

\section{Discussion and conclusion} 

With the data at hand it is not yet possible to reveal the nature of the unseen companion in BPS\,BS\,16981$-$0016. It might be either a massive WD or a NS. Although a triple scenario is still possible, we consider the stripping of the sdB progenitor by stable RLOF to the massive companion as the more likely formation scenario, because it matches quite well with theoretical models. In this case, the companion should be spun up considerably (Wu et al. \cite{wu18}; Benvenuto et al. \cite{benvenuto15}) and might be observable as a pulsar in the radio or the X-ray domain. The Very Large Array Sky Survey (VLASS, Gordon et al. \cite{gordon20}) and the Master X-ray Catalogue (XRAY\footnote{https://heasarc.gsfc.nasa.gov/W3Browse/all/xray.html}) do not contain any measurements at the coordinates of BPS\,BS\,16981$-$0016.

Due to their long orbital periods and rather small RV variations, those binaries are hard to detect by RV variability surveys. About one third of the apparently single sdBs, which are quite challenging to explain in the context of binary evolution (Geier et al. \cite{geier22}). Some of them might be yet undetected objects of this kind and stable RLOF to compact objects a more important formation channel for sdBs than predicted. 

If some of those objects should be SN\,Ia progenitors, this new type of binary stars would also contribute to the observed SN\,Ia rate. However, it is still unclear whether this contribution would be significant. An entirely new class of binaries with NS companions on the other hand would also be a major discovery, because it would not only significantly increase the number of NS as such, but in particular the number of NS with accurately determined masses. 

\begin{acknowledgements}  

We thank Andreas Irrgang for the development of the spectrum and SED- fitting tools and his contributions to the model atmosphere grids. We also thank Max Pritzkuleit for sharing data with us. 

V.S. and H.D. are supported by the Deutsche Forschungsgemeinschaft (DFG) through grants GE2506/12-1 and GE2506/17-1, respectively.

This work presents results from the European Space Agency (ESA) space mission Gaia. Gaia data are being processed by the Gaia Data Processing and Analysis Consortium (DPAC). Funding for the DPAC is provided by national institutions, in particular the institutions participating in the Gaia MultiLateral Agreement (MLA). The Gaia mission website is https://www.cosmos.esa.int/gaia. The Gaia archive website is https://archives.esac.esa.int/gaia.

Based on observations with the Isaac Newton Telescope operated by the Isaac Newton Group at the Observatorio del Roque de los Muchachos of the Instituto de Astrofisica de Canarias on the island of La Palma, Spain.

This paper makes use of data obtained from the Isaac Newton Group Archive which is maintained as part of the CASU Astronomical Data Centre at the Institute of Astronomy, Cambridge.

This paper includes data collected by the TESS mission, which are publicly available from the Mikulski Archive for Space Telescopes (MAST). Funding for the TESS mission is provided by NASA’s Science Mission directorate.

Based on observations obtained with the Samuel Oschin Tele-
scope 48-inch and the 60-inch Telescope at the Palomar Observatory
as part of the Zwicky Transient Facility project. ZTF is supported
by the National Science Foundation under Grant No. AST-1440341
and a collaboration including Caltech, IPAC, the Weizmann Institute
for Science, the Oskar Klein Center at Stockholm University, the
University of Maryland, the University of Washington, Deutsches
Elektronen-Synchrotron and Humboldt University, Los Alamos Na-
tional Laboratories, the TANGO Consortium of Taiwan, the Univer-
sity of Wisconsin at Milwaukee, and Lawrence Berkeley National
Laboratories. Operations are conducted by COO, IPAC, and UW.

This research has made use of the VizieR catalogue access tool, CDS, Strasbourg, France (DOI:\,10.26093/cds/vizier). The original description 
 of the VizieR service was published in 2000, A\&AS 143, 23
                                                   
\end{acknowledgements}

\newpage
\appendix
\section{Tables and figures}

\begin{figure*}[t!]
\begin{center}
	\resizebox{10cm}{!}{\includegraphics[angle=90]{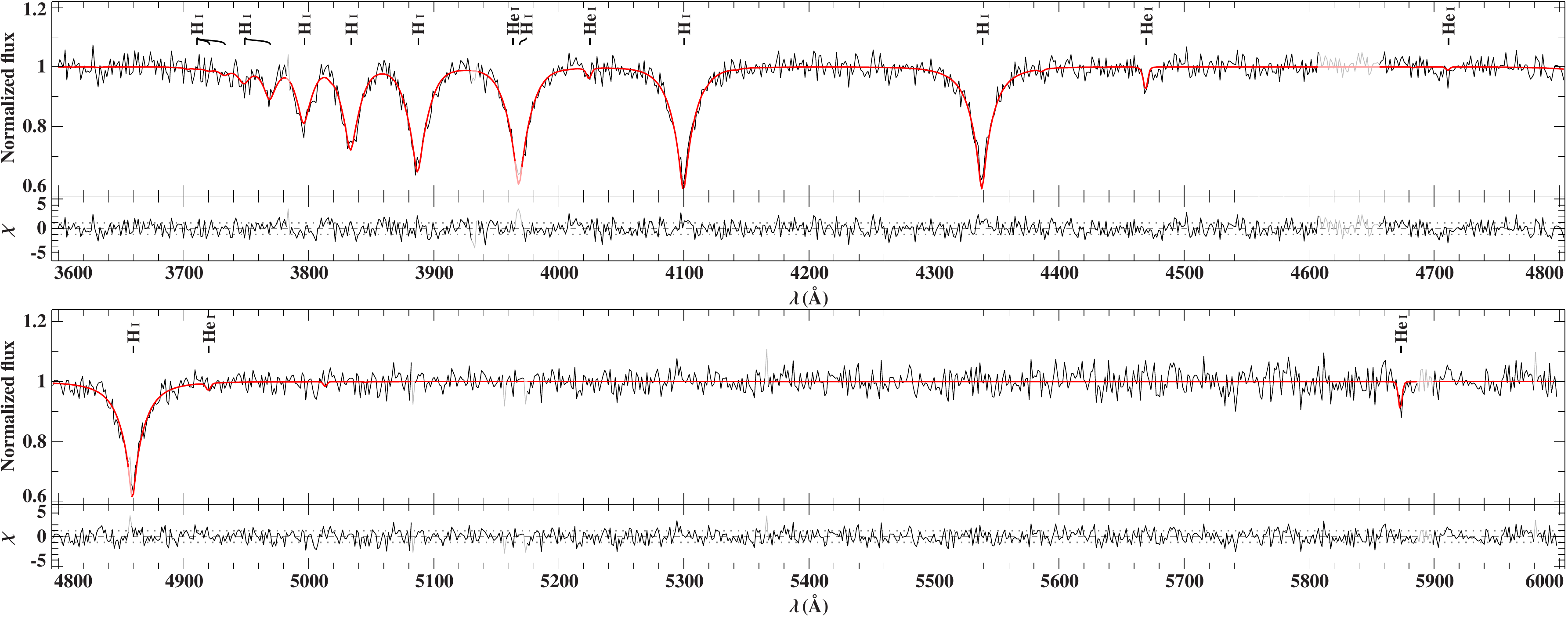}}
\end{center} 
\caption{Comparison of the best-fitting model spectrum (red line) to the INT spectrum taken in 2010 (solid black line). The residuals are shown in the respective bottom panels. The dashed horizontal lines mark deviations in terms of $\pm1\,\sigma$, i.e. values of $\chi=\pm1$. }
\label{spectro_fit_2010}
\end{figure*}

\begin{table*}
\caption{Parameters of BPS\,BS\,16981$-$0016}
\label{tab:params}
\begin{center}
\begin{tabular}{ll} \hline\hline
\noalign{\smallskip}
Name                                                     &  BPS\,BS\,16981$-$0016  \\
                                                         &  Gaia\, DR3\,3649963989549165440 \\
\noalign{\smallskip}                                     
\hline                                                   
\noalign{\smallskip}                                     
Right ascension (J2000) $\alpha$                         &  14:33:30.9                   \\
Declination (J2000) $\delta$                             &  -01:14:43.0                  \\
Magnitude in G-band                                      &  $14.300\pm0.003\,{\rm mag}$  \\
Parallax $\varpi$                                        &  $1.37\pm0.05\,{\rm mas}$ \\
Distance $d$                                             &  $730_{-27}^{+28}\,{\rm pc}$ \\
\noalign{\smallskip}                                     
\hline                                                   
\noalign{\smallskip}                                     
Effective temperature sdB $T_{\rm eff}$                  &  $29560 \pm 270$ \\
Surface gravity sdB $\log{g}$                            &  $5.88\pm0.05$ \\
Helium abundance sdB $\log{n({\rm He)}/n({\rm H)}}$      &  $-2.66\pm0.13$ \\
\noalign{\smallskip}                         
\hline                                       
\noalign{\smallskip}
Color excess E(44-55)                                    &  $0.085\pm0.002\,{\rm mag}$ \\
Angular diameter $\log{\Theta\,{\rm [rad]}}$             &  $11.049_{-0.010}^{+0.011}$ \\
Radius sdB $R_{\rm sdB}$                                 &  $0.145\pm0.007\,R_{\rm \odot}$ \\
Mass sdB $M_{\rm sdB}$                                   &  $0.58_{-0.15}^{+0.20}\,M_{\rm \odot}$ \\
Luminosity sdB $L_{\rm sdB}$                             &  $14.5_{-2.1}^{+2.4}\,L_{\rm \odot}$ \\
\noalign{\smallskip}
\hline
\noalign{\smallskip}
Orbital period $P$                                       &  $892.5\pm60.2\,{\rm d}$  \\
Orbital eccentricity $e$                                 &  $0.36\pm0.14$ \\
Orbital separation$^{\rm *}$ $a$                         &  $2.269\pm0.056\,{\rm mas}$ \\
Orbital inclination$^{\rm *}$ $i$                        &  $56.8\pm3.2^{\circ}$ \\
Astrometric mass function $m_{\rm f}$                    &  $0.78\pm0.15\,M_{\rm \odot}$ \\
Companion mass $M_{\rm comp}$                            &  $1.50_{-0.45}^{+0.37}\,M_{\rm \odot}$ \\
\noalign{\smallskip}
\hline
\end{tabular}
\tablefoot{$^{\rm *}$Calculated following Halbwachs et al. (\cite{halbwachs22}) by CDS and provided as additional columns to the Gaia DR3 catalogue Part 3: Non-single stars (I/357).}
\end{center}
\end{table*}

\begin{table}
\caption{Radial velocities of BPS\,BS\,16981$-$0016}
\label{tab:rvs}
\begin{center}
\begin{tabular}{ll} 
\hline\hline
\noalign{\smallskip}
mid-HJD       &  RV                    \\
              &  [${\rm km\,s^{-1}}$]  \\
\noalign{\smallskip}                     
\hline
\noalign{\smallskip}
2455377.50234   &   -69 $\pm$ 7   \\
2459797.38698   &   -21 $\pm$ 6   \\
2459797.39753   &   -13 $\pm$ 7   \\
2459798.37475   &   -30 $\pm$ 7   \\
2459798.38530   &   -31 $\pm$ 7   \\
2459799.38686   &   -22 $\pm$ 6   \\
2459800.38605   &   -25 $\pm$ 6   \\
2459800.39661   &   -9  $\pm$ 6   \\
2459800.40719   &   -3  $\pm$ 6   \\
2459801.37819   &   -12 $\pm$ 10  \\
2459801.38878   &   -17 $\pm$ 10  \\
2459801.39990   &   -19 $\pm$ 10  \\
2459802.37648   &   -31 $\pm$ 7   \\
2459802.38813   &   -32 $\pm$ 6   \\
2459803.36143   &   -23 $\pm$ 7   \\
2459803.37198   &   -19 $\pm$ 6   \\
2459803.38254   &   -25 $\pm$ 6   \\
\noalign{\smallskip}
\hline
\end{tabular}
\end{center}
\end{table}

\begin{figure*}[t!]
\begin{center}
	\resizebox{14cm}{!}{\includegraphics{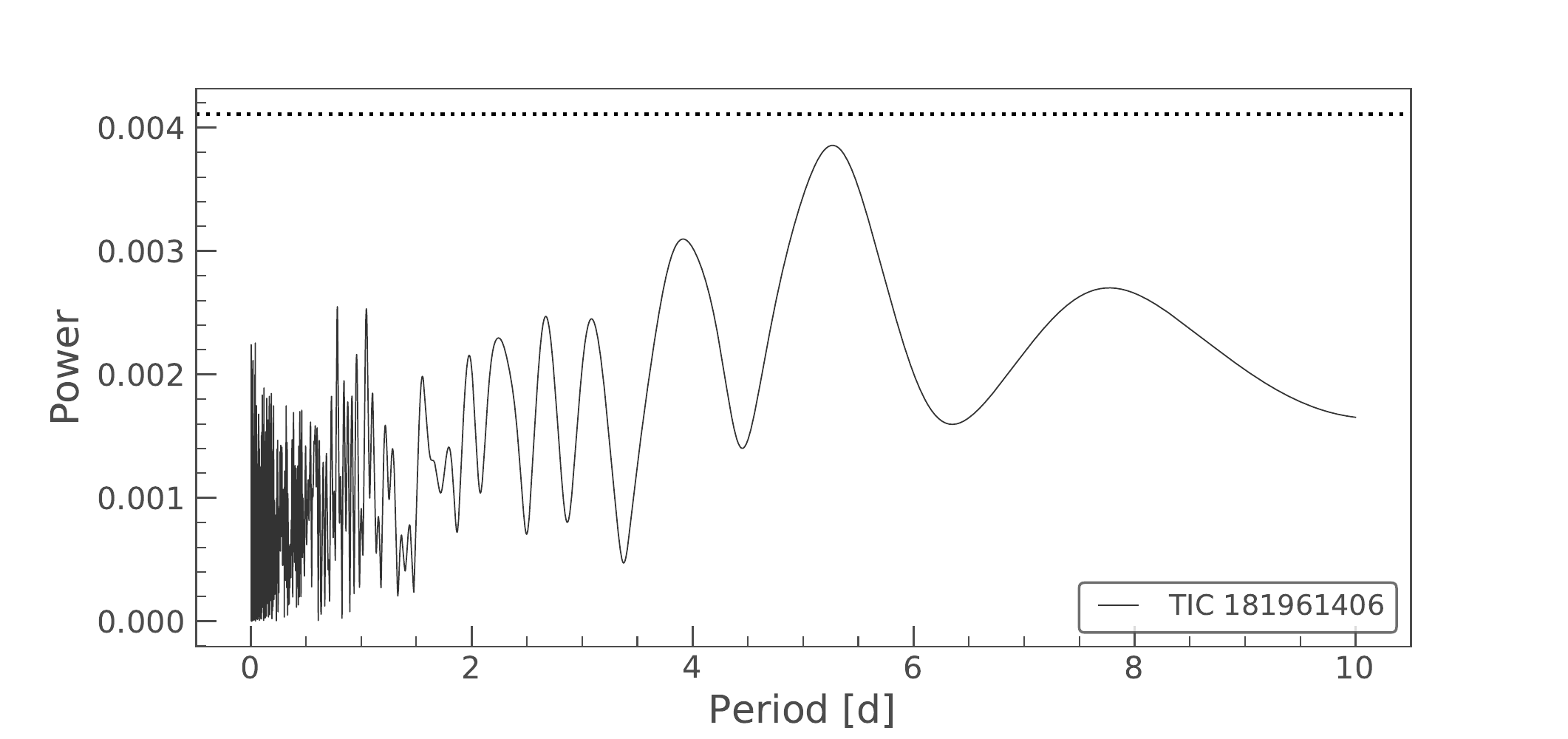}}
	\resizebox{14cm}{!}{\includegraphics{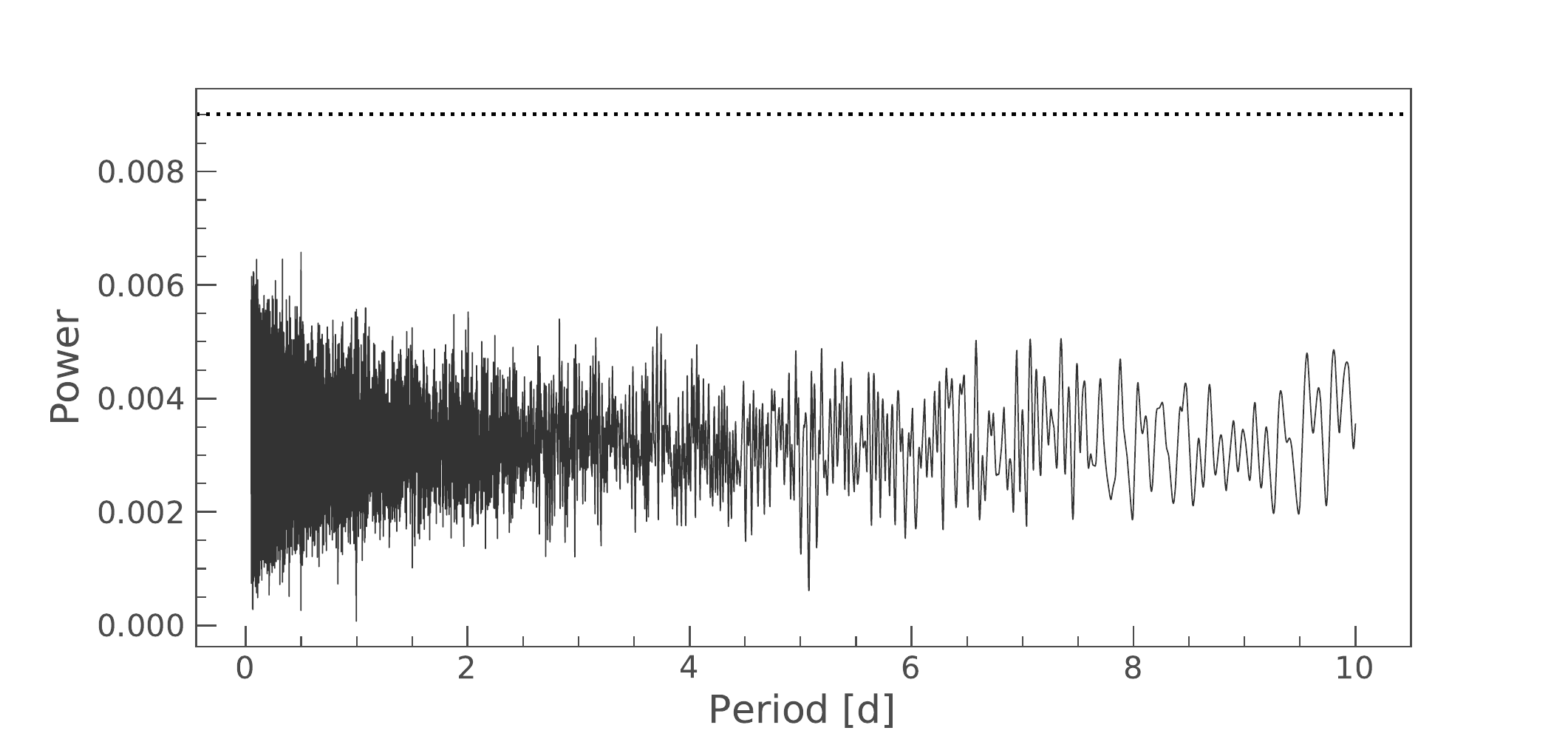}}
	\resizebox{14cm}{!}{\includegraphics{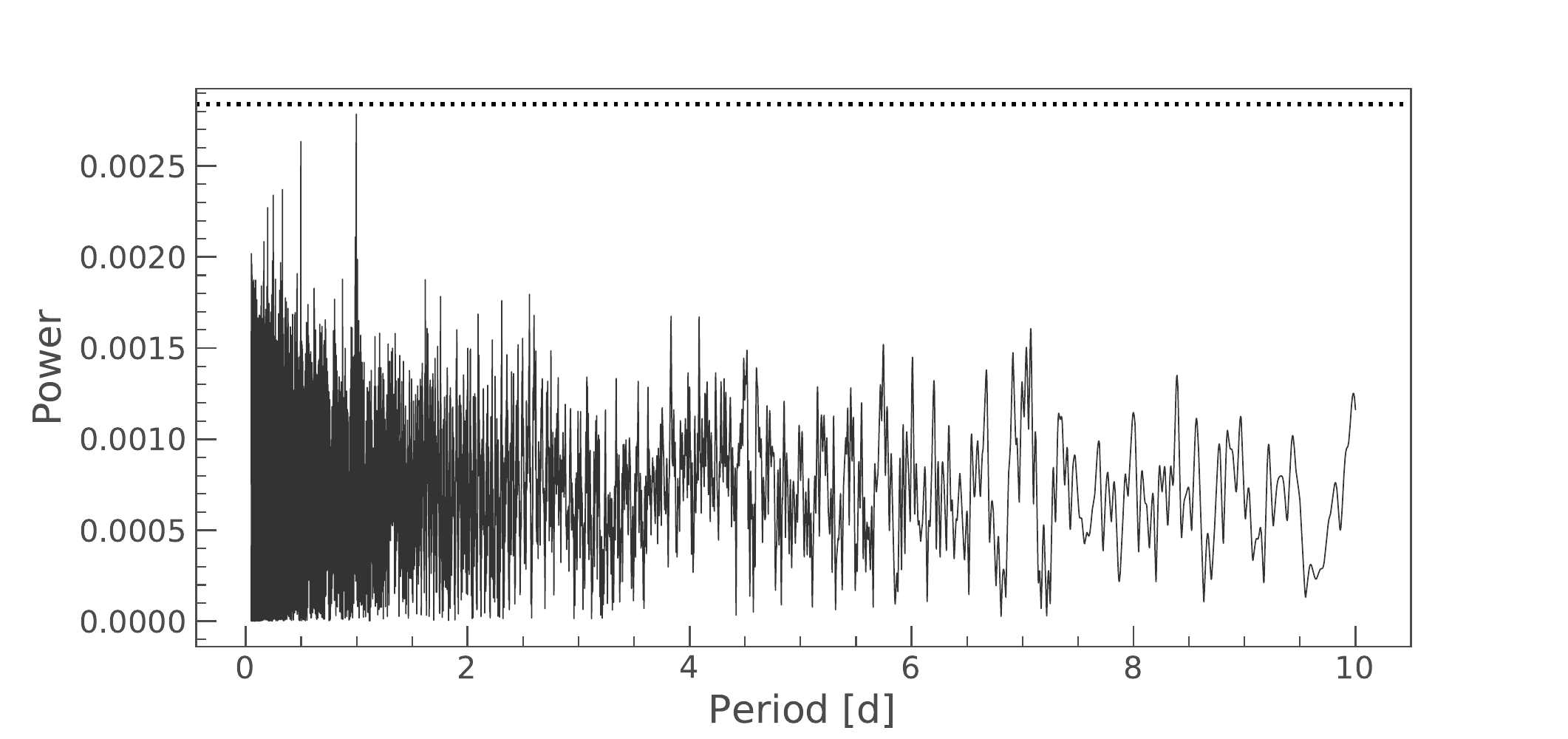}}
\end{center} 
\caption{Power spectra of light curves of BPS\,BS\,16981$-$0016 (TIC\,181961406) from TESS (upper panel) as well as ZTF g-band (middle panel) and r-band (lower panel). The $1\%$ false-alarm level is indicated by the dashed line.}
\label{lc}
\end{figure*}

                            
\end{document}